# Tunable Quantum Anomalous Hall Effects in Ferromagnetic van der Waals Heterostructures

Feng Xue[1,2], Yusheng Hou[3], Zhe Wang[4], Zhiming Xu[2], Ke He[1,2,5], Ruqian Wu[6], Yong Xu[2,5,7,8, *], Wenhui Duan[1,2,5,9, *]

[1] Beijing Institute of Quantum Information Science, Beijing 100193, China

[2] State Key Laboratory of Low-Dimensional Quantum Physics, Department of Physics, Tsinghua University, Beijing 100084, China

[3] Guangdong Provincial Key Laboratory of Magnetoelectric Physics and Devices, Center for Neutron Science and Technology, School of Physics, Sun Yat-Sen University, Guangzhou, 510275, China

[4] State Key Laboratory of Surface Physics and Key Laboratory for Computational Physical Sciences (MOE) and Department of Physics, Fudan University, Shanghai 200433, China

[5] Frontier Science Center for Quantum Information, Beijing 100084, China

[6] Department of Physics and Astronomy, University of California, Irvine, CA 92697-4575, USA

[7] Tencent Quantum Laboratory, Tencent, Shenzhen, Guangdong 518057, China

[8] RIKEN Center for Emergent Matter Science (CEMS), Wako, Saitama 351-0198, Japan

[9] Institute for Advanced Study, Tsinghua University, Beijing 100084, China

**ABSTRACT**

The quantum anomalous Hall effect (QAHE) has unique advantages in topotronic applications, but it is still challenging to realize the QAHE with tunable magnetic and topological properties for building functional devices. Through systematic first-principles calculations, we predict that the in-plane magnetization induced QAHE with Chern numbers $C = \pm 1$ and the out-of-plane magnetization induced QAHE with high Chern numbers $C = \pm 3$ can be realized in a single material candidate, which is composed of van der Waals (vdW) coupled Bi and $MnBi_2Te_4$ monolayers. The switching between different phases of QAHE can be controllable by multiple ways, such as applying strain or (weak) magnetic field or twisting the vdW materials. The prediction of an experimentally available material system hosting robust, highly tunable

QAHE will stimulate great research interest in the field. Our work opens a new avenue for the realization of tunable QAHE and provides a practical material platform for the development of topological electronics.

* Corresponding authors: duanw@tsinghua.edu.cn; yongxu@mail.tsinghua.edu.cn;

**INTRODUCTION**

The quantum anomalous Hall effect (QAHE), featured by a quantized Hall conductance at zero magnetic field and the topologically protected chiral edge states, has been widely studied in recent years [1-3]. Typically, an out-of-plane magnetization is needed to break the time reversal symmetry and all in-plane mirror symmetries. To this end, most prior theoretical and experimental work have been pursuing QAHE in materials with an out-of-plane magnetization (OPM-QAHE) [4-12]. Nevertheless, QAHE can also be induced by an in-plane magnetization, i.e., IPM-QAHE, as revealed by Liu *et al*. in 2013 based on two-dimensional (2D) point group symmetry analysis [13]. It is known that most magnetic films prefer in-plane magnetic anisotropy as the thickness is reduced to a few monolayers [14-17], so it is more appealing to establish QAHE in geometries with in-plane magnetization for their integration in nanodevices. The main hurdle for the realization of IPM-QAHE is that the in-plane magnetization does not necessarily break all symmetries that can forbid the anomalous Hall conductance in most materials and this research area hence has not been much explored [13,18-22]. Furthermore, the search for topological materials with high and tunable Chern numbers is another attractive interdisciplinary fundamental topic as the multiple dissipationless edge conduction channels may significantly improve the performance of devices [23]. It is believed that materials with high-Chern-numbers lead to new topological phases with exotic elementary excitations [24]. Therefore, finding innovative material platforms for experimental realization of IPM-QAHE, preferably also with multiple quantum edge states, is crucial for the advance of topotronic physics and applications.

Given that research of van der Waals (vdW) magnets are rapidly advancing [25-29] and their magnetization can be conveniently controlled by bias or strain [30], it is intriguing to explore QAHE in ferromagnetic (FM) vdW heterostructures. Unlike many magnetic insulators with unwanted hybridization and charge transfer at their interfaces with topological insulators (TIs), 2D vdW magnets safeguard the topological surface states (TSSs) and it appears to be an excellent strategy to delve into heterostructures with different FM vdW films and TIs for the realization of QAHE at a reasonably high temperature. What's more, it is also intriguing and meaningful to explore if IPM-QAHE and OPM-QAHE may coexist and switchable in these heterostructures. This is expected to greatly enrich the pool of topological quantum materials and provide new ideas for the design of multifunctional topological devices.

In this work, we theoretically explore the electronic, magnetic and topological properties of the vdW heterostructure of ferromagnetic $MnBi_2Te_4$ septuple layer (SL) and nonmagnetic Bi bilayer (BL), as depicted in Fig. 1(a). Through first-principles calculations, we demonstrate that IPM-QAHE with Chern numbers $C = \pm1$ as well as OPM-QAHE with Chern numbers $C = \pm3$ can be realized in the $Bi/MnBi_2Te_4$ heterostructure. As a strain of 2-3% can flip the magnetization from in-plane to out-plane, topological phase transition between the two QAHE states is easily accessible. Moreover, the system may switch between $C = +3$ and $C = -3$ phases by simply twisting the FM layers in the heterostructure by 60° without need of an external magnetic field to reverse the spin direction. The controllable magnetism and topological phase transition in this ferromagnetic vdW heterostructure are illustrated in Fig. 1(b). This work provides a unique material platform for the realization and manipulation of multiple QAHE phases as required in tunable nanodevices.

## RESULTS
### Structural and magnetic properties

The bulk Bi and $MnBi_2Te_4$ crystalize in the same rhombohedral layered structure with the space group $R\bar{3}m$, and the layers stack along the perpendicular direction with an ABC sequence. The $MnBi_2Te_4$ monolayer consists of a Te-Bi-Te-Mn-Te-Bi-Te SL, wherein Mn atoms are arranged in a triangular lattice with spins parallel to each other. The single layer of Bi in a honeycomb lattice has a buckled structure, with the nearest atoms displacing in the perpendicular direction, and thus it is usually called BL in the literature. Bi BL is a topological insulator with the band gap of 0.47 eV by either GGA+U or HSE06 calculations [Fig. S1a (S1c)]. The $MnBi_2Te_4$ SL is a FM semiconductor with a band gap of 0.19 eV (0.56 eV) by GGA+U (HSE06) calculations [Fig. S1b (S1d)]. The optimized lattice constants of the $MnBi_2Te_4$ SL and Bi BL with the inclusion of the spin-orbit coupling (SOC) effect are 4.37 and 4.38 Å, respectively, consistent with previous theoretical results [31, 32]. Such a good lattice match is advantageous for experimental synthesis of Bi/MnBi$_2$Te$_4$ heterostructure.

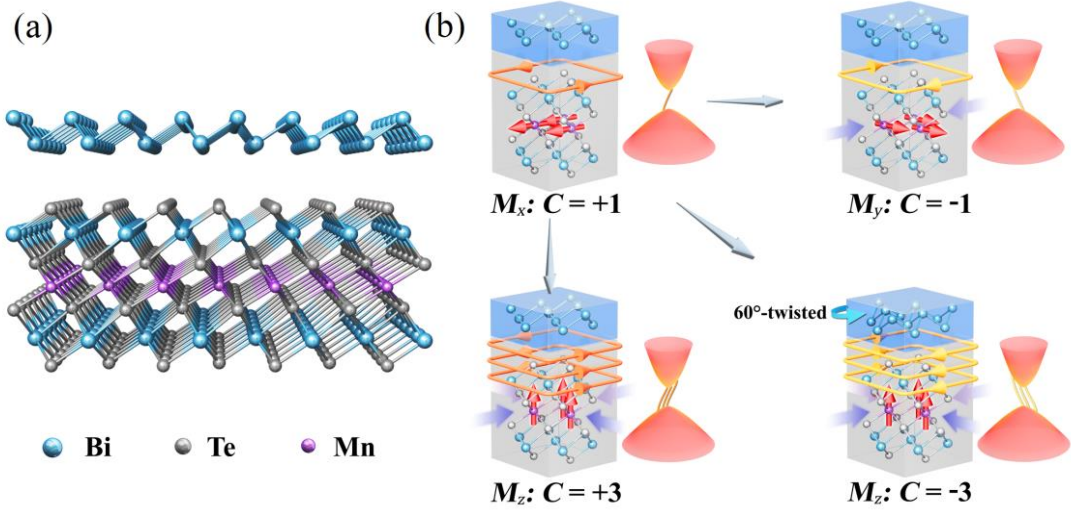

FIG 1. (a) Crystal structure of Bi/MnBi$_2$Te$_4$ vdW heterostructure. (b) Schematic diagram showing controllable magnetism and topological phase transition in the material. The magnetization orientation ($M_i$: $i = x, y, z$; red arrows) is tunable by applying strain (denoted by purple arrows) or magnetic field. QAH states with controllable Chern numbers are realized by changing the magnetization orientation or twisting the vdW heterostructure.

There are six possible high-symmetry alignments between MnBi$_2$Te$_4$ SL and Bi BL in the Bi/MnBi$_2$Te$_4$ heterostructure. As shown in Fig. S2, three of them have the inner Bi atoms of Bi BL sitting above the hollow, Te and Bi sites of MnBi$_2$Te$_4$ (denoted as P1, P2 and P3, respectively). The other three alignments are obtained by twisting the Bi BL by 60° (denoted as P1′, P2′ and P3′). According to their binding energies in Table S2, the most stable configuration is the P1 structure depicted in Figs. 1(a) and S2(a). The optimized interlayer distance between MnBi$_2$Te$_4$ SL and Bi BL is 2.84 Å, a value which is comparable to that (2.73 Å) between adjacent SLs in bulk MnBi$_2$Te$_4$ [33], indicating that the interaction between MnBi$_2$Te$_4$ SL and Bi BL is also of the vdW type. Indeed, the calculated binding energy ($E_b$) is only -0.402 eV/unit cell for the most preferred P1 structure. Here, the binding energy is define as $E_b = E_{Bi/MnBi_2Te_4} - E_{Bi} - E_{MBT}$, with $E_{Bi}$, $E_{MnBi_2Te_4}$ and $E_{Bi/MnBi_2Te_4}$ representing the energies of isolated Bi BL, MnBi$_2$Te$_4$ SL and Bi/MnBi$_2$Te$_4$ heterostructure, respectively. Similar to the free-standing MnBi$_2$Te$_4$ SL, the exchange interaction between the nearest neighbor Mn$^{2+}$ ions in Bi/MnBi$_2$Te$_4$ is FM and its value is 1.08 meV (Table S1). This indicates that MnBi$_2$Te$_4$ retains its FM ground state in Bi/MnBi$_2$Te$_4$. Noticeably, the magnetic easy axis of Bi/MnBi$_2$Te$_4$ is changed to in-plane, with a tiny magnetic anisotropy energy (MAE) of $E_x - E_z = -0.15$ meV/Mn (Fig. S3). This is different from the out-of-plane magnetic easy axis of the free-standing MnBi$_2$Te$_4$ SL. Further calculations show that the energy barrier of spin rotation within the *x-y* plane is negligibly small (Fig. S3), indicating that the Bi/MnBi$_2$Te$_4$ is a typical 2D XY magnets. In such easy-plane 2D systems, the magnetic order can be stabilized by the finite size effect as has been demonstrated recently in 1T-VSe$_2$ and CrCl$_3$ monolayers [17,34].

**In-plane magnetization induced QAHE**

The in-plane magnetic Bi/MnBi$_2$Te$_4$ heterostructure provides an interesting platform for exploring the IPM-QAHE. As known, all in-plane mirror symmetries must be broken for the realization of QAHE since they forbid the anomalous Hall conductance. While this symmetry breaking can be naturally achieved in geometries with an out-of-plane magnetization, the circumstance needs to be carefully checked for IPM-QAHE geometries. In principle, an in-plane magnetization ***m*** cannot break the mirror plane $M_m$ perpendicular to ***m***. The present Bi/MnBi$_2$Te$_4$ system has three mirror planes $M_y$, $M_1$ and $M_2$, which are related to each other via the $C_{3z}$ rotations as displayed in Fig. 2(e). Hence, the QAHE is allowed only when the in-plane magnetization is not perpendicular to any of them.

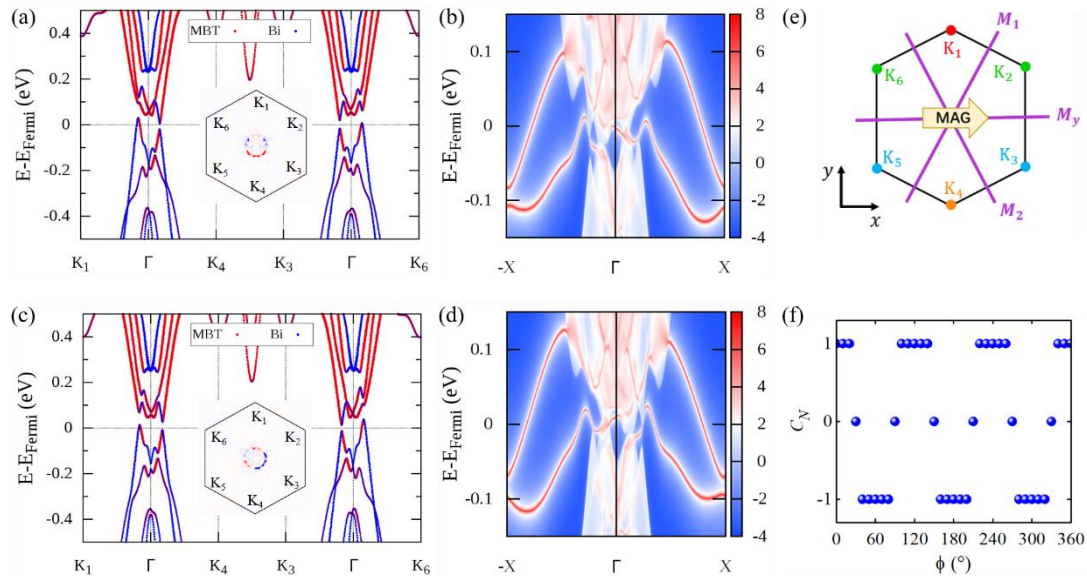

*FIG 2. (a) and (c) DFT calculated band structures along the path of $K_1 \to \Gamma \to K_4 \to K_3 \to \Gamma \to K_6$ for in-plane magnetization along $\phi = 0°$ and $\phi = 60°$, respectively. The red and blue lines represent the components of $MnBi_2Te_4$ and Bi, respectively. Inset shows the distribution of Berry curvature in the first BZ. (b) and (d) The edge states corresponding to (a) and (c), respectively. (e) The first BZ of $Bi/MnBi_2Te_4$. The three violet lines denote the three mirror planes $M_y$, $M_1$ and $M_2$. The magnetization is aligned to the x direction. The six K points are marked by dots. Same colors indicate that they are related by the $M_y \otimes \mathcal{T}$ symmetry operation. (f) The Chern number as a function of the in-plane magnetization direction, where $\phi$ is the angle between the magnetization and x direction.*

To fulfil this requirement, we align the magnetization of $Bi/MnBi_2Te_4$ heterostructure along the *x* direction (parallel to the $M_y$ plane) and calculate its band structure and topological properties. As shown in Fig. 2(e), when the magnetization is along the *x* direction ($\phi = 0°$), there are four nonequivalent K points in the first Brillouin zone (BZ), denoted as $K_1$, $K_4$, $K_3$ and $K_6$. Owing to the preserved $M_y \otimes \mathcal{T}$ operation, $K_2$ and $K_5$ are equivalent to $K_6$ and $K_3$, respectively. Figure 2(a) shows the component-resolved band structure along the path of $K_1 \to \Gamma \to K_4 \to K_3 \to \Gamma \to K_6$. One may see that the bands near the gap have strong intermixing between $MnBi_2Te_4$ and Bi. Although there is no global band gap, the electron and hole pockets are very tiny, which is also supported by calculations with the HSE06 functional (Fig. S4). The distribution of Berry curvature in the first BZ is shown in the inset of Fig. 2(a), which yields a Chern number of $C = +1$. Clearly, this means that $Bi/MnBi_2Te_4$ can offer an opportunity of realizing the IPM-QAHE. This is further confirmed by the existence of one chiral edge state that connects the conduction and valence bands of a $Bi/MnBi_2Te_4$ ribbon [Fig. 2(b)]. By rotating the magnetization orientation to $\phi = 60°$, the band structure, shown in Fig. 2(c), looks the same as that of $\phi = 0°$, except that the paths are reversed. Consequently, the sign of Berry curvature around $\Gamma$ point in the first BZ changes from positive (Inset of Fig. 2a) to negative (Inset of Fig. 2c), giving a negative Chern number of $C = -1$ and one chiral edge state propagating in the opposite direction (Fig. 2d). As the magnetic energy in the *x-y* plane is nearly isotropic (Fig. S3), the magnetization can be rotated in the base plane by applying a tiny magnetic field or uniaxial strain.

In addition, we calculate the Chern number of Bi/MnBi$_2$Te$_4$ as a function of the in-plane magnetization direction, and show the results in Fig. 2(f) and Fig. S5. As expected, when the in-plane magnetization is perpendicular to one of the three mirror planes (i.e., $\phi = 30° + n \times 60°$, with $n$ = 0, 1, 2, 3, 4, 5), the Chern number drops to zero due to the preservation of mirror symmetries (Figs. S5b and S5f). Furthermore, the Chern number exhibits a periodic jumping between $C = +1$ and $C = -1$ with an interval of 60° when we rotate the magnetization in the lateral plane (Fig. S5), which is a common feature of the in-plane magnetization-induced QAHE [13,19,20,22]. The opposite Chern numbers with an interval of 60° can be explained by imposing a time reversal operation $\mathcal{T}$ and then a 120° clockwise rotation on $\phi$.

**Out-of-plane magnetization induced QAHE**

For device applications, it is more interesting to see if the IPM-QAHE of the Bi/MnBi$_2$Te$_4$ heterostructure can be further tuned by simple approaches, such as external magnetic fields, strain and interlayer twisting. In section A we have shown that the Bi/MnBi$_2$Te$_4$ has a tiny MAE of -0.15 meV/Mn, indicating its magnetization can be easily switched between in-plane and out-of-plane states by applying a perpendicular magnetic field. In addition, as the interaction between the MnBi$_2$Te$_4$ and Bi layers is of weak vdW type, one may twist them to a desired angle. We explore the effect of these two factors separately below.

When the spin is aligned to the $z$ direction by applying an out-of-plane magnetic field, there are only two nonequivalent K points in the first BZ, namely K$_1$ and K$_4$. For the most stable P1 configuration of Bi/MnBi$_2$Te$_4$ with an out-of-plane magnetization, the band structure along the path of $K_1 \rightarrow \Gamma \rightarrow K_4 \rightarrow M \rightarrow \Gamma$ is plotted in Fig. 3(a). Compared with bands of in-plane magnetization in Fig. 2(a), a noticeable change is the appearance of a global direct band gap. We then examine the topological properties by calculating the distribution of Berry curvature in the first BZ and the anomalous Hall conductivity (AHC) $\sigma_{xy} = C\frac{e^2}{h}$ ($h$ is the Planck constant and $e$ is the elementary charge). As illustrated in Fig. 3(b), the out-of-plane magnetization leads to a high Chern number of $C = +3$ in the band gap, with large Berry curvature values around the $\Gamma$ point. In Fig.

3(c), three chiral edge states can be clearly observed for one-dimensional Bi/MnBi$_2$Te$_4$ nanoribbon, further demonstrating it is tuned to the OPM-QAHE phase with a high Chern number $C = +3$.

We next turn to the effect of twisting angles. By twisting the P1 configuration of Bi/MnBi$_2$Te$_4$ by 60°, we arrive at the P1′ configuration as shown in Fig. S2(d). The main difference between these two structures is the position of the outmost atoms of Bi BL. For the P1 configuration, the outmost Bi atoms sit above the interfacial Bi atoms. While for the P1′ configuration, they move to the top of interfacial Te atoms. Such a structural change leads to a significant modification of the band structure around the Fermi level [Figs. 3(a) and 3(d)], resulting in a sign change of the Berry curvature around the Γ point [Insets of Figs. 3(b) and 3(e)]. The integration of Berry curvature over the whole BZ gives a Chern number of $C = -3$, which again is confirmed by the result of quantized AHC $\sigma_{xy}$ with $C = -3$ [Fig. 3(e)] in the band gap, as well as the presence of three chiral gapless edge modes from the calculation for a ribbon [Fig. 3(f)]. Herein, a 3% (-3%) strain is applied to P1 (P1′) configuration to clearly show the plateau of anomalous Hall conductivity (AHC), which does not change their intrinsic topological properties.

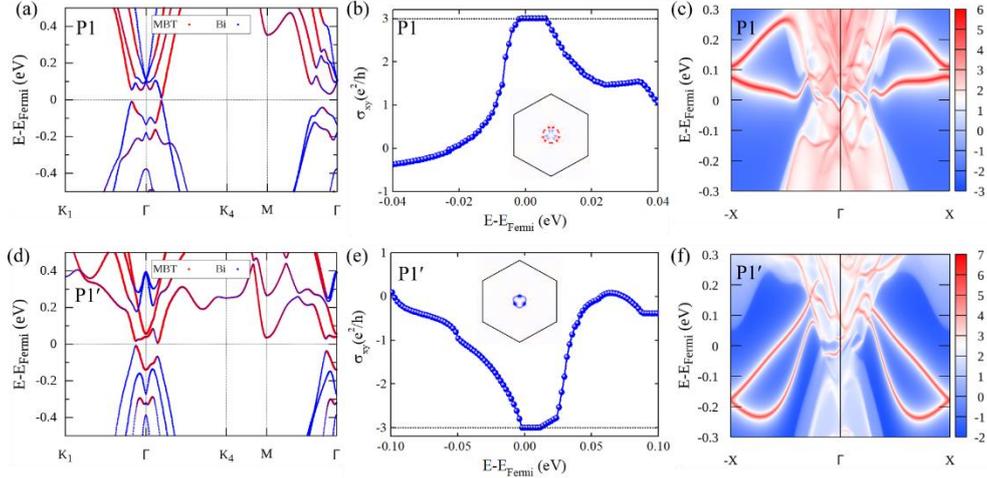

FIG 3. *Topological properties of P1 (P1′) configuration of Bi/MnBi$_2$Te$_4$ with an out-of-plane magnetization under a 3% (−3%) strain. (a) DFT calculated DFT band structure for P1 configuration. (b) AHC as a function of the Fermi energy for P1 configuration, where the distribution of Berry Curvature in the first BZ is shown in the inset. (c) The*

*chiral edge states of one-dimensional Bi/MnBi$_2$Te$_4$ nanoribbon with P1 configuration. (d)-(f) same as (a)-(c) but for P1′ configuration.*

**Strain effects**

Beside the external magnetic field and twisting angle, applying an external strain is also an effective route to tune the material properties. We impose a biaxial strain ranging from -5% and 5% on both the P1 and P1′ configurations of Bi/MnBi$_2$Te$_4$, and examine their relative energies, magnetic properties and topological features. As shown in Fig. 4, under a compressive strain, the magnetic easy axes of both P1 and P1′ configurations are switched from in-plane to out-of-plane; meanwhile, the most stable alignment configuration changes from P1 to P1′. Note that, the topological properties as described in Fig. 3 are unchanged under a large range of strain, -4% ~ 5%, as long as we maintain the out-of-plane magnetization. Therefore, Bi/MnBi$_2$Te$_4$ not only displays IPM-QAHE with $C = \pm1$, but also can be tuned to have robust OPM-QAHE with a high Chern number $C = \pm3$. As these phases are switchable by means of biaxial strain or magnetic field, Bi/MnBi$_2$Te$_4$ is an ideal platform to realize tunable QAHE, similar to the NiAsO$_3$ and PdSbO$_3$ monolayers proposed recently by Li *et al*. [21].

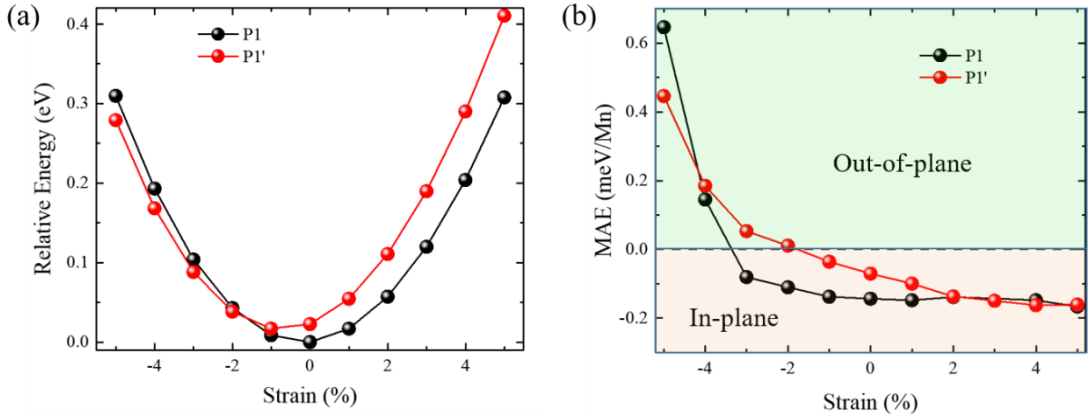

*FIG 4. Strain effects on the relative energy and magnetic anisotropy energy (MAE) of P1 and P1′ structures. (a) The relative energy and (b) MAE as a function of biaxial strain. The energy of P1 structure without strain is set to zero as a reference.*

**DISCUSSION AND CONCLUSION**

Considering the vdW nature of MnBi$_2$Te$_4$ SL and Bi BL and their successful fabrications in experiments [35,36], we believe that the vdW Bi/MnBi$_2$Te$_4$ heterostructure is feasible for experimental synthesis [37]. On the other hand, ever since the bulk MnBi$_2$Te$_4$ was theoretically identified as an intrinsic magnetic topological insulator [12,38,39], the OPM-QAHE has been extensively examined in MnBi$_2$Te$_4$ films, particularly for those thicker than 5 SLs [35,40]. A single MnBi$_2$Te$_4$ SL is a topologically trivial 2D FM semiconductor [12,39]. In this regard, it is of interest to realize QAHE in its heterostructure built with Bi BL and our finding of multiple QAHE phases in Bi/MnBi$_2$Te$_4$ with such a thin thickness should be greatly beneficial for the design of ultra-thin topological spintronic devices [41].

In summary, based on systematic first-principles calculations, we demonstrate the highly tunable QAHE in a vdW heterostructure of 2D FM semiconductor MnBi$_2$Te$_4$ SL and nonmagnetic semiconductor Bi BL. We find that the Bi/MnBi$_2$Te$_4$ heterostructure not only displays the unusual in-plane magnetization induced QAHE with a Chern number of $C = \pm 1$, but also has tunable QAHE with high Chern numbers of $C = \pm 3$ under out-of-plane magnetic fields. The interlayer twisting or biaxial strain may provide more means for the control of its topological phases. Specifically, the topological quantum phase transition from $C = \pm 1$ to $C = \pm 3$ and $C = +3$ to $C = -3$ can be realized in a single system by applying strain or perpendicular magnetic field. Our findings greatly extend the practical routes for searching materials with the highly tunable QAHE and should spur more experimental and theoretical explorations in this realm.

**METHODS**

As shown in Fig. 1(a), the Bi/MnBi$_2$Te$_4$ heterostructure is modeled by a slab with a Bi BL on a single MnBi$_2$Te$_4$ SL layer. The lattice constant in the lateral plane was fixed to the optimized size of Bi BL, while the vertical positions of all atoms are relaxed. A vacuum space of 15 Å between adjacent slabs was set to eliminate the spurious interactions between periodic images. Density functional theory (DFT) calculations were performed by using the projector augmented wave [42,43] method as implemented in the Vienna *ab initio* simulation package (VASP) [44]. The exchange-correlation interaction was described by using the functional proposed by Perdew, Burke, and Ernzerhof (PBE) [45]. The semi-core states of Mn 3$p$ and Bi 5$d$ were treated as valence electrons. The cutoff energy of the plane-wave basis set was chosen to be 400 eV. A Γ-centered 15 × 15 × 1 Monkhorst-Pack k-point mesh was adopted for structural relaxation, and a denser 23 × 23 × 1 mesh was used for calculating their magnetic anisotropy energies. The structural relaxations were performed using the conjugate gradient method with criteria that require the force acting on each atom smaller than 0.001 eV/Å and energy change less than $10^{-6}$ eV. The strong electron correlation effect for the localized orbitals of Mn was treated by the DFT+$U$ method [46] using $U$ = 4 eV. The vdW corrections were invoked through the DFT-D3 method [47] in all calculations. In addition, the Heyd-Scuseria-Ernzerhof (HSE) hybrid functional was used [48] in test calculations. The edge state calculations were performed using the Wannier Tools [49], where the maximally localized Wannier functions were constructed by using the software package Wannier90 [50] interfaced with VASP.

**FUNDING**

This work was supported by the Ministry of Science and Technology of China (Grant Nos. 2018YFA0307100 and 2018YFA0305603), the Basic Science Center Project of NSFC (Grant No. 51788104), the National Science Fund for Distinguished Young Scholars (Grant No. 12025405), the National Natural Science Foundation of China (Grant Nos. NSFC-12104518 and 11874035), the Beijing Advanced Innovation Center for Future Chip, and the Beijing Advanced Innovation Center for Materials Genome Engineering. Work at the University of California, Irvine was supported by DOE-BES of USA (Grant No. DE-FG02-05ER46237). Computer simulations were partially performed at the U.S. Department of Energy Supercomputer Facility (NERSC).